\documentclass[aps,prl,reprint,groupedaddress]{revtex4-1}

\usepackage{graphicx}
\usepackage{amsmath}
\usepackage{hyperref}

\newcommand{\bs}[1]{\textbf{\textsl{#1}}}
\newcommand{\abs}[1]{\left\lvert #1 \right\rvert}
\newcommand{\chppm}{\left(\frac{c}{\hbar} \: m_{\gamma \prime} \right)}

\newcommand{\oiint}{\mathop{{\int\!\!\!\!\int}\mkern-22mu \bigcirc}}

\begin{document}

\title{Microwave cavity hidden sector photon threshold crossing}

\author{Rhys G. Povey}
\email[]{rhys.povey@uwa.edu.au}
\author{John G. Hartnett}
\author{Michael E. Tobar}
\affiliation{School of Physics, University of Western Australia, WA 6009 Australia}

\date{\today}

\begin{abstract}
Hidden sector photons are a weakly interacting slim particle arising from an additional U(1) gauge symmetry predicted by many standard model extensions. We present and demonstrate a new experimental method using a single microwave cavity to search for hidden sector photons. Only photons with a great enough energy are able to oscillate into hidden sector photons of a particular mass. If our cavity is driven on resonance and tuned over the corresponding threshold frequency, there is an observable drop in the circulating power signifying the creation of hidden sector photons. This approach avoids the problems of microwave leakage and frequency matching inherent in photon regeneration techniques.
\end{abstract}

\pacs{14.80.-j}

\maketitle

Extensions to the standard model of particle physics commonly predict an extra hidden U(1) gauge symmetry~\cite{abel,goodsell}. This corresponds to a hidden sector of particles where by the only interaction to standard matter is through kinetic mixing between the hidden sector photon $\gamma'$ and ordinary photon $\gamma$~\cite{okun,holdom}. The mass of the hidden sector photon is unknown but thought to be very light and belong to a class of hypothetical particles known as WISPs (Weakly Interacting Slim/Sub-eV Particles)~\cite{jaeckel_lowenergy}. Typically, experiments to search for the hidden sector photon in this sub-eV range have utilized the (resonantly enhanced) photon regeneration ``light shining through a wall'' (LSW) method~\cite{okun,redondo_lsw} with either a laboratory~\cite{axionexp2,Hoogeveen1,axionexp3,axionexp4,jaeckel_cavity} or astronomical~\cite{axionastro1,Hoogeveen2} source of particles. Much progress has been made with these experiments using laser light~\cite{BFRT1,BFRT2,BMV,ahlers_paraphoton,GammeV,ahlers_laser,afanasev_2008,fouche,afanasev_2009,ALPS,ALPS2010} and, more recently, microwaves~\cite{rp_lsw1,admx_lsw} to place limits on the existence of hidden sector photons.

In this letter we propose and experimentally test a novel method for hidden sector photon searches. Differing from the standard approach of photon regeneration our design uses only a single microwave cavity and indirectly looks for hidden sector photons by energy leaving the system. A measurable drop in circulating power can potentially be observed if the cavity is tuned over the hidden sector photon creation threshold frequency. This approach does not suffer from the problems of microwave leakage and frequency matching inherent in traditional LSW experiments.

If the hidden sector photon has a mass then only photons with an energy greater than the hidden sector photon rest energy can oscillate into hidden sector photons, i.e. $h f \geq m_{\gamma\prime}\,c^2$ for $\gamma \rightarrow \gamma^\prime$. Thus if we tune the frequency of a cavity, and hence energy of the resonating photons, over the threshold value then we can expect to see a sudden drop in circulating power as some photons start suddenly oscillating into hidden sector photons and escaping the cavity. This allows for a sensitive detection of a hidden sector photon event but only for the narrow range of masses which the cavity can tune through.

The threshold crossing effect is determined by calculating the loss due to photons leaving the cavity through photon-hidden sector photon oscillations. Our model is based upon the construction of a hidden sector photon quality factor
\begin{align}
Q_{\gamma\prime} = \omega_\gamma \frac{\textmd{time average stored energy in cavity}}{\textmd{power loss to hidden sector photons}} \label{eq:HSPQdefn}.
\end{align}
We derive an expression for $Q_{\gamma\prime}$ in a vacuum by first determining the stress-energy-momentum tensor.

The Lagrangian density describing the two fields is~\cite{jaeckel_cavity}
\begin{multline}
\mathcal{L} = -\frac{1}{4} F^{\mu\,\nu}\,F_{\mu\,\nu} -\frac{1}{4} B^{\mu\,\nu}\,B_{\mu\,\nu} -\frac{1}{2} \chi\,F^{\mu\,\nu}\,B_{\mu\,\nu} \\ +\frac{1}{2} \chppm^2\,B^{\mu}\,B_{\mu}, \label{eq:Lagrangian}
\end{multline}
where $F^{\mu\,\nu}$ is the standard electromagnetic field strength tensor for gauge field $A^{\mu}$, $B^{\mu\,\nu}$ is the hidden sector field strength tensor for gauge field $B^{\mu}$, $\chi$ is the hidden sector mixing parameter and $m_{\gamma\prime}$ is the hidden sector photon mass.
Following the Belinfante construction~\cite{belinfante}, the time-averaged stress-energy-momentum tensor for complex gauge fields is
\begin{multline}
\langle T^{\mu\,\nu} \rangle = \frac{1}{2} \Re \Biggl( -F^{\mu\lambda}\,{F^{\nu}_{\;\;\lambda}}^* -B^{\mu\lambda}\,{B^{\nu}_{\;\;\lambda}}^* -\chi\,F^{\mu\lambda}\,{B^{\nu}_{\;\;\lambda}}^* \\-\chi\,B^{\mu\lambda}\,{F^{\nu}_{\;\;\lambda}}^* +\chppm^2 B^{\mu}\,{B^{\nu}}^* -\eta^{\mu\nu}\biggl( -\frac{1}{4}F^{\alpha\beta}\,{F_{\alpha\beta}}^* \\-\frac{1}{4}B^{\alpha\beta}\,{B_{\alpha\beta}}^* -\frac{1}{2}\chi\,F^{\alpha\beta}\,{B_{\alpha\beta}}^* +\frac{1}{2}\chppm^2 B^{\alpha}{B_{\alpha}}^* \biggr) \Biggr) .
\end{multline}
We choose a temporal Lorenz Coulomb gauge for $A^{\mu}$ such that $A^0=0$, $\partial_{\mu}A^{\mu}=0$ and $B^0=0$.

By shifting the Lagrangian $B^\mu \rightarrow \tilde{B}^\mu - \chi A^\mu$ and solving the equations of motions, the hidden sector gauge field generated by a resonating vacuum cavity is~\cite{jaeckel_cavity}
\begin{align*}
\tilde{B}^j (\bs{b},t) = \chi \chppm^2 \iiint\limits_{V} \frac{e^{i \: k_{\gamma \prime} \abs{\bs{b}-\bs{a}}}}{4\pi \abs{\bs{b}-\bs{a}}} A^j(\bs{a},t) \: \mathrm{d}^3\bs{a} ,
\end{align*}
where $V$ is the cavity volume, $A^j = (i\,\omega_\gamma\,\sqrt{\mu})^{-1}\,E^j$ is the electromagnetic gauge field and $k_{\gamma\prime}$ is the hidden sector photon wave number such that 
\begin{align*}
k_{\gamma\prime}=\sqrt{\left(\frac{\omega_\gamma}{c}\right)^2 - \chppm^2} ,
\end{align*}
for small $\chi$.
We assume that $A^j$ is generated in a purely standard sense and is found by solving Maxwell's equations within the cavity (see Ref.~\cite{rp_lsw1} for field equations in a cylindrical cavity). For outgoing waves $A^j(\bs{a},t)=A^j(\bs{a})\,e^{-\,i\,\omega_\gamma\,t}$ and $B^j(\bs{b},t)=B^j(\bs{b})\,e^{-\,i\,\omega_\gamma\,t}$.
The integral for $B^j$ can not be dealt with analytically but we can make a good approximation by considering $B^j$ at an arbitrarily large distance from the cavity, such that $\abs{\bs{b}} \gg \abs{\bs{a}}$, and near the threshold frequency, where $\lambda_{\gamma\prime} \gg \abs{\bs{a}}$. Thus we consider a Taylor series expansion about $\abs{\bs{a}}=0$.

The energy flow outside the cavity ($A^{\mu}=0$) from hidden sector photons is
\begin{align*}
\langle {S_{\gamma\prime}}^{j} \rangle = \langle T^{j\,0} \rangle c \, \vert_{A^{\mu}=0} = \frac{c}{2}\Re\left(-B^{j\lambda}\,{B^{0}_{\;\;\lambda}}^*\right) .
\end{align*}
To obtain the total power loss we integrate the flux over a surface enclosing the cavity, most conveniently a sphere of arbitrarily large radius $s$, 
\begin{align*}
P_{\gamma\prime}=& \oiint\limits_{\abs{\bs{b}}=s} \langle \bs{S}_{\gamma \prime}(\bs{b}) \rangle \cdot \mathrm{d}^2 \bs{b} .
\end{align*}

For the energy stored in the cavity we consider only the standard photons ($B^{\mu}=0$). The time-averaged energy density is
\begin{align*}
u = \langle T^{0\,0} \rangle \vert_{B^{\mu}=0} = \frac{1}{2}\Re\left(-F^{0\,\lambda}\,{F^{0}_{\;\;\lambda}}^*+\frac{1}{4}F^{\alpha\beta}\,{F_{\alpha\beta}}^*\right)
\end{align*}
and the total energy, $U$, is the integral of $u$ over the cavity.

\begin{figure}[t]
\includegraphics[width=0.45\textwidth]{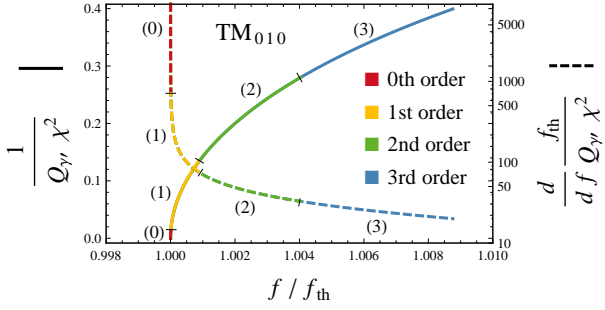}
\caption{(Color online). Plots of the 0th, 1st, 2nd and 3rd order approximations of $1/Q_{\gamma\prime}$ (solid line) and its derivative (dashed line) against frequency as a proportion of the threshold frequency. These are for the $\mathrm{TM}_{\,0\,1\,0}$ mode in a cylindrical cavity of unity aspect ratio ($L=2R$) with tuning performed by changing the size of the cavity whilst maintaining the aspect ratio. Each approximation is shown to the limit of its validity determined by $\left(k_{\gamma\prime} \abs{\bs{a}}\right)^{q+1} < 0.01$ where $q$ is the order.\label{fig:invQ1}}
\end{figure}

Returning to equation \eqref{eq:HSPQdefn}, $Q_{\gamma\prime}^{\;-1}=\omega_\gamma^{\;-1} \, P_{\gamma \prime} / U $. In a cylindrical cavity the hidden sector photon quality factor is minimized by the $\mathrm{TM}_{\,0\,1\,0}$ mode. For a cavity of length $L$, radius $R$, and resonance frequency $\omega_\gamma = \varsigma_{0,1} \: c / R$, where $\varsigma_{0,1}$ is the first root of the Bessel $J_0$ function, the 0th order approximation of the hidden sector photon quality factor is
\begin{align*}
\frac{1}{Q_{\gamma\prime}} \stackrel{0}{=} \frac{2\:\chi^2\,L}{3} \left(\frac{c}{\omega_\gamma}\right)^4 \chppm^4 \sqrt{\left(\frac{\omega_\gamma}{c}\right)^2 - \chppm^2} .
\end{align*}

\begin{figure}[t]
\includegraphics[width=0.45\textwidth]{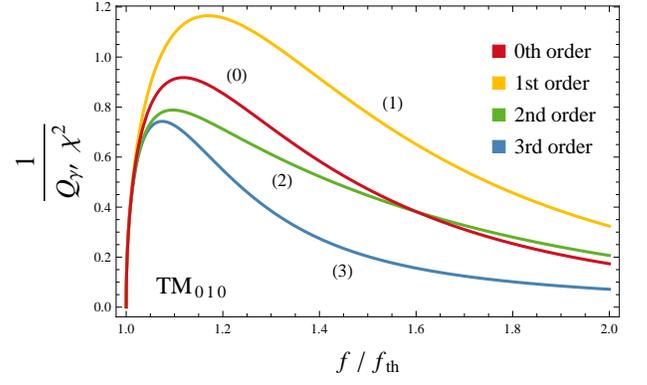}
\caption{(Color online). Plots of $1/Q_{\gamma\prime}$ approximations, as labeled, over longer frequency ranges well outside their valid regions.\label{fig:invQ2}}
\end{figure}

Plots of $1/Q_{\gamma\prime}$ for the $\mathrm{TM}_{\,0\,1\,0}$ mode, in a cylindrical cavity of unity aspect ratio ($L=2R$), at 0th, 1st, 2nd and 3rd order approximations are given in Fig.~\ref{fig:invQ1}. The frequency tuning is assumed to be by uniformly changing the size of the cavity whilst maintaining the aspect ratio. The validity region of the approximations may appear quite small but it's only this small region we're interested in. In a microwave cavity around $10\:\mathrm{GHz}$ the third order approximation is valid up to $100\:\mathrm{MHz}$ from the threshold.
The hidden sector photon loss does not continue to increase at higher frequencies. As shown in Fig.~\ref{fig:invQ2} all of the $1/Q_{\gamma\prime}$ approximations predict a turning point and decline back towards zero.

We can now relate the hidden sector photon quality factor to observables in a microwave cavity experiment.
For a cavity with two ports we are able to measure the incident power, transmitted power and reflected power.
Losses within the cavity will manifest themselves as changes to the coupling.
When on resonance the power terms are related to the input, $\beta_1$, and output, $\beta_2$, coupling parameters by
\begin{align}
P_\mathrm{refl} &= \frac{(1-\beta_1+\beta_2)^2}{(1+\beta_1+\beta_2)^2} \; P_\mathrm{inc} ,\\
P_\mathrm{trans} &= \frac{4\,\beta_1\,\beta_2}{(1+\beta_1+\beta_2)^2} \; P_\mathrm{inc} \label{eq:Ptrans},\\
P_\mathrm{diss} &= \frac{4\,\beta_1}{(1+\beta_1+\beta_2)^2} \; P_\mathrm{inc} .
\end{align}
The quality factor relationships are
\begin{align*}
\frac{1}{Q_L} = \frac{1+\beta_1+\beta_2}{Q_0} , &&
\frac{1}{Q_0} = \frac{1}{Q_{\gamma}} + \frac{1}{Q_{\gamma\prime}} ,
\end{align*}
where $Q_L$ is the loaded quality factor and $Q_0$ is the unloaded quality factor idealizing an isolated cavity with no input or output. $Q_\gamma$ and $Q_{\gamma\prime}$ represent the standard photon and hidden sector photon portions of $Q_0$ respectively.

Searching for hidden sector photons is then carried out by taking measurements from our cavity related to $Q_{\gamma\prime}$ as the frequency is tuned. To continue driving the cavity on resonance a frequency lock system is required.

\begin{figure}[tb]
\includegraphics[width=0.45\textwidth]{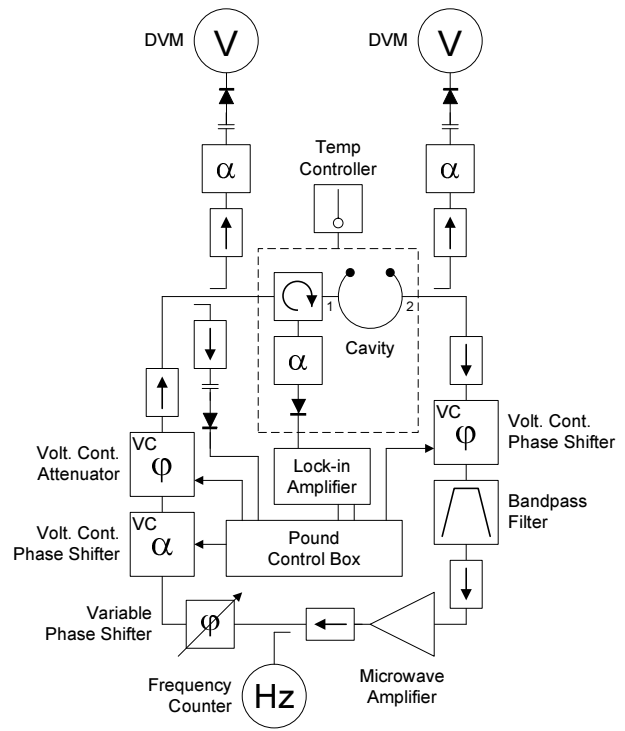}
\caption{Microwave circuit of our prototype experiment.\label{fig:experiment}}
\end{figure}

To demonstrate this threshold crossing approach we set up a simple prototype experiment. The essential elements are a tunable microwave cavity, a locked driving signal and frequency and power measurements. For our prototype we used a copper cavity at room temperature in a vacuum. The length and diameter of the cavity were approximately $2\:\mathrm{c\,m}$, and the $\mathrm{TM}_{\,0\,1\,0}$ resonance frequency was around $10.402\:\mathrm{GHz}$.
Tuning was carried out by a heating element and temperature controller between $300\:\mathrm{K}$ and $324\:\mathrm{K}$, corresponding to a $4\:\mathrm{MHz}$ frequency span.
To drive and lock to the cavity a Pound control scheme loop oscillator was used~\cite{Pound}.
A schematic of the experiment setup is given in Fig.~\ref{fig:experiment}.

In our prototype experiment we applied only a very small amount of tuning and thus the external quality factors remain constant, $\beta_1/Q_0=\eta_1$ and $\beta_2/Q_0=\eta_2$. An initial measurement of $\beta_1$, $\beta_2$ and $Q_L$ was used to determine $\eta_1$ and $\eta_2$, these are shown in Table~\ref{tab:meas}.

\begin{table}
\caption{Initial measurements.\label{tab:meas}}
\begin{ruledtabular}
\begin{tabular}{ccc|crclclcc}
&& $Q_L$ && 2000 &$\pm$& 200 &&&&\\
&& $\beta_1$ && 0.4146 &$\pm$& 0.0083 &&&& \\
&& $\beta_2$ && 0.0570 &$\pm$& 0.0078 &&&& \\
&& $\eta_1$ &(& 0.1409 &$\pm$& 0.0169 &)& $\times 10^{-3}$ &&\\
&& $\eta_2$ &(& 0.0194 &$\pm$& 0.0046 &)& $\times 10^{-3}$ &&\\
\end{tabular}
\end{ruledtabular}
\end{table}

From equations \eqref{eq:Ptrans} it follows that
\begin{align}
\frac{P_\mathrm{trans}}{P_\mathrm{inc}} &= \frac{4}{\left( \frac{1}{\sqrt{\eta_1\,\eta_2}} \frac{1}{Q_\gamma}+\frac{1}{\sqrt{\eta_1\,\eta_2}} \frac{1}{Q_{\gamma\prime}}+\sqrt{\frac{\beta_1}{\beta_2}}+\sqrt{\frac{\beta_2}{\beta_1}} \right)^2} \nonumber\\
&= \frac{4}{\bigl( X(f)+Y(f)+B \bigr)^2} \label{eq:Pratio1},
\end{align}
where $B$ is a constant, $X$ is some function of frequency and $Y=(1/\sqrt{\eta_1\,\eta_2})(1/Q_{\gamma\prime})$ also a function of frequency.

Searching for a hidden sector photon event can be made easier by looking at the derivative of equation \eqref{eq:Pratio1} where there is a large spike at the threshold. As $1/Q_{\gamma\prime}$ is extremely small we can use a 0th order approximation of the Taylor series about $Y=0$. This gives us
\begin{align}
\frac{\mathrm{d}}{\mathrm{d} f} \frac{P_\mathrm{trans}}{P_\mathrm{inc}} \stackrel{0}{=} \frac{-4 \bigl( X'(f) + Y'(f) \bigr)}{\bigl( X(f) + B \bigr)^3} \label{eq:Pratio1d}.
\end{align}

Given some data for $P_\mathrm{trans}/P_\mathrm{inc}$ we can test for a particular threshold crossing frequency $f_\mathrm{th}=m_{\gamma\prime}\,c^2/h$. Below the threshold $1/Q_{\gamma\prime}=0$ and thus we can create a fit to the data
\begin{align*}
\frac{P_\mathrm{trans}}{P_\mathrm{inc}} \stackrel{f<f_\mathrm{th}}{=} \frac{4}{\left( X(f)+B \right)^2} = \xi(f)
\end{align*}
Now we can look at the data above the threshold and take the residuals of the derivative, equation \eqref{eq:Pratio1d},
\begin{align}
\mathrm{Residual}\left( \frac{\mathrm{d}}{\mathrm{d} f} \frac{P_\mathrm{trans}}{P_\mathrm{inc}} \right) \stackrel{0}{=} \frac{- \bigl( \xi(f) \bigr)^{3/2}}{\sqrt{\eta_1\,\eta_2}} \frac{\mathrm{d}}{\mathrm{d} f} \frac{1}{Q_{\gamma\prime}}(f) \label{eq:smalltuneres},
\end{align}
which is proportional to $\chi^2$.
The comparison with data will depend on the standard deviation of the derivative of $P_\mathrm{trans}/P_\mathrm{inc}$, i.e. of the $\xi'(f)$ fit, and the frequency resolution (bin width) of the data.

Data from our experiment was collected in 4 continuous runs, 2 of heating up and 2 of cooling down. Detector voltages were converted into the incident and transmitted power levels as a function of frequency. Each data set of $P_\mathrm{trans}/P_\mathrm{inc}$ was then put into frequency bins of $\Delta f = $ $100$, $1000$ and $10 000$ $\mathrm{Hz}$. These binned data sets were then searched for hidden sector photons by scanning through and testing for a threshold frequency in each bin. The central difference quotient of the residuals were compared against the worst possible discretization and certainty of equation \eqref{eq:smalltuneres}. For our analysis a 3rd order approximation of $1/Q_{\gamma\prime}$ was used.

\begin{figure}[tb]
\includegraphics[width=0.45\textwidth]{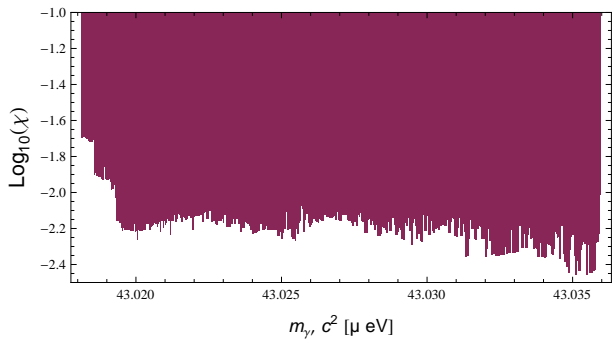}
\caption{Hidden sector photon exclusion plot from the prototype threshold crossing experiment.\label{fig:result1a}}
\end{figure}

Limits of $\chi$ were placed with a one standard deviation, $\sigma$, buffer, calculated from the $\xi'(f)$ fit, added to the data. The compiled results are given in Fig.~\ref{fig:result1a}. The limit covers only a narrow range of hidden sector photon masses related to the narrow range of frequencies tuned across. As a prototype experiment the results on $\chi$ are not competitive with other experimental tests. However, there is significant room for improvement and we will show the possibility of a competitive experiment.

In the case of no hidden sector photons the limit on $\chi$ is related to $\sigma$ and the discretized equation \eqref{eq:smalltuneres} as
\begin{align*}
\chi \propto \sqrt{\sigma} \; \sqrt{\frac{1}{Q_\gamma}} \; \sqrt[4]{\Delta f} \; \sqrt[4]{f_\mathrm{th}} \label{eq:chiprop}.
\end{align*}
To improve our limit each of these items needs to be minimized.

As expected the limit has a strong dependence on the cavity quality factor $Q_\gamma$. Using superconducting Niobium cavities at cryogenic temperatures $Q_\gamma \sim 10^{10}$ is possible~\cite{Q10}.

The standard deviation of our difference quotient data, $\sigma$ is proportional to the deviation in the original data which in turn is proportional to the amplitude noise of our system. By using better microwave amplifiers, detectors and measuring techniques this can easily be improved. Of particular issue in our prototype experiment, the lock-in amplifier was not ideal for a low $Q$ cavity. A process of tuning, stabilizing, then measuring will produce much better results (albeit taking longer) than continuously drifting while measuring. Our experiment had an amplitude noise of $-80\,\mathrm{dBc}/\mathrm{Hz}$ whilst values around $-120\,\mathrm{dBc}/\mathrm{Hz}$ are obtainable~\cite{eugene1}.

The minimal usable bin width $\Delta f$ is determined by the best obtainable resolution which is limited by the frequency stability according to
\begin{gather*}
\textmd{Resolution} = \textmd{Bandwidth} \times \textmd{Line splitting factor} ,\\
\textmd{Bandwidth} = f / Q_L ,\\
\textmd{Line splitting factor} = Q_L \times \sigma_y(\tau) ,
\end{gather*}
where $\sigma_y$ is the frequency stability of the oscillator and $\tau$ is the sampling time.
Any frequency drift needs to be less than $\textmd{resolution}/\textmd{sampling time}$. A line splitting factor $\sim 10^{-7}$ is potentially achievable~\cite{dick_lsf} so for a cavity of quality factor $\sim 10^{10}$ at $10\:\mathrm{GHz}$ the smallest usable bin width is $\Delta f \sim 10^{-6} \,\mathrm{Hz}$.

Lower frequency thresholds of investigation, i.e. larger cavity sizes, also slightly improve results but this is not something we want to strictly minimize. Comparing the above achievable values to our prototype experiment we find that a limit of $\chi<10^{-10}$ is possible.

Furthermore there is potential to improve the limits on $\chi$ by employing more advanced data analysis techniques. In data with improved stability it may be possible to use optimal filter analysis to get a tighter hidden sector photon restriction.

In this letter we have proposed and formulated a new experimental method to search for hidden sector photons using microwave cavities. A prototype experiment was carried out to demonstrate the idea. With an improved cryogenic experimental setup this approach allows exceptionally small values of the hidden sector photon mixing parameter $\chi$ to be probed over a narrow range of hidden sector photon masses without the necessity of a detection cavity, in which microwave leakage and resonance matching are problematic. The achievable results are well beyond the current limits and are in potential discovery space.

The authors would like to thank Joerg Jaeckel and Ian McArthur for helpful discussions. This work was supported by the Australian Research Council grant number DP1092690.


\begin{thebibliography}{31}%
\makeatletter
\providecommand \@ifxundefined [1]{%
 \@ifx{#1\undefined}
}%
\providecommand \@ifnum [1]{%
 \ifnum #1\expandafter \@firstoftwo
 \else \expandafter \@secondoftwo
 \fi
}%
\providecommand \@ifx [1]{%
 \ifx #1\expandafter \@firstoftwo
 \else \expandafter \@secondoftwo
 \fi
}%
\providecommand \natexlab [1]{#1}%
\providecommand \enquote  [1]{``#1''}%
\providecommand \bibnamefont  [1]{#1}%
\providecommand \bibfnamefont [1]{#1}%
\providecommand \citenamefont [1]{#1}%
\providecommand \href@noop [0]{\@secondoftwo}%
\providecommand \href [0]{\begingroup \@sanitize@url \@href}%
\providecommand \@href[1]{\@@startlink{#1}\@@href}%
\providecommand \@@href[1]{\endgroup#1\@@endlink}%
\providecommand \@sanitize@url [0]{\catcode `\\12\catcode `\$12\catcode
  `\&12\catcode `\#12\catcode `\^12\catcode `\_12\catcode `\%12\relax}%
\providecommand \@@startlink[1]{}%
\providecommand \@@endlink[0]{}%
\providecommand \url  [0]{\begingroup\@sanitize@url \@url }%
\providecommand \@url [1]{\endgroup\@href {#1}{\urlprefix }}%
\providecommand \urlprefix  [0]{URL }%
\providecommand \Eprint [0]{\href }%
\@ifxundefined \urlstyle {%
  \providecommand \doi  [0]{\begingroup \@sanitize@url \@doi}%
  \providecommand \@doi [1]{\endgroup \@@startlink {\doibase
  #1}doi:\discretionary {}{}{}#1\@@endlink }%
}{%
  \providecommand \doi  [0]{doi:\discretionary{}{}{}\begingroup
  \urlstyle{rm}\Url }%
}%
\providecommand \doibase [0]{http://dx.doi.org/}%
\providecommand \Doi [0]{\begingroup \@sanitize@url \@Doi }%
\providecommand \@Doi  [1]{\endgroup\@@startlink{\doibase#1}\@@Doi}%
\providecommand \@@Doi [1]{#1\@@endlink}%
\providecommand \selectlanguage [0]{\@gobble}%
\providecommand \bibinfo  [0]{\@secondoftwo}%
\providecommand \bibfield  [0]{\@secondoftwo}%
\providecommand \translation [1]{[#1]}%
\providecommand \BibitemOpen [0]{}%
\providecommand \bibitemStop [0]{}%
\providecommand \bibitemNoStop [0]{.\EOS\space}%
\providecommand \EOS [0]{\spacefactor3000\relax}%
\providecommand \BibitemShut  [1]{\csname bibitem#1\endcsname}%
\bibitem [{\citenamefont {Abel}\ \emph {et~al.}(2008)\citenamefont {Abel},
  \citenamefont {Goodsell}, \citenamefont {Jaeckel}, \citenamefont {Khoze},\
  and\ \citenamefont {Ringwald}}]{abel}%
  \BibitemOpen
  \bibfield  {author} {\bibinfo {author} {\bibfnamefont {S.~A.}\ \bibnamefont
  {Abel}}, \bibinfo {author} {\bibfnamefont {M.~D.}\ \bibnamefont {Goodsell}},
  \bibinfo {author} {\bibfnamefont {J.}~\bibnamefont {Jaeckel}}, \bibinfo
  {author} {\bibfnamefont {V.~V.}\ \bibnamefont {Khoze}}, \ and\ \bibinfo
  {author} {\bibfnamefont {A.}~\bibnamefont {Ringwald}},\ }\Doi
  {10.1088/1126-6708/2008/07/124} {\bibfield  {journal} {\bibinfo  {journal}
  {JHEP},\ }\textbf {\bibinfo {volume} {07}},\ \bibinfo {pages} {124} (\bibinfo
  {year} {2008})},\ \Eprint {http://arxiv.org/abs/0803.1449} {arXiv:0803.1449
  [hep-ph]} \BibitemShut {NoStop}%
\bibitem [{\citenamefont {Goodsell}\ \emph {et~al.}(2009)\citenamefont
  {Goodsell}, \citenamefont {Jaeckel}, \citenamefont {Redondo},\ and\
  \citenamefont {Ringwald}}]{goodsell}%
  \BibitemOpen
  \bibfield  {author} {\bibinfo {author} {\bibfnamefont {M.}~\bibnamefont
  {Goodsell}}, \bibinfo {author} {\bibfnamefont {J.}~\bibnamefont {Jaeckel}},
  \bibinfo {author} {\bibfnamefont {J.}~\bibnamefont {Redondo}}, \ and\
  \bibinfo {author} {\bibfnamefont {A.}~\bibnamefont {Ringwald}},\ }\Doi
  {10.1088/1126-6708/2009/11/027} {\bibfield  {journal} {\bibinfo  {journal}
  {JHEP},\ }\textbf {\bibinfo {volume} {11}},\ \bibinfo {pages} {027} (\bibinfo
  {year} {2009})},\ \Eprint {http://arxiv.org/abs/0909.0515} {arXiv:0909.0515
  [hep-ph]} \BibitemShut {NoStop}%
\bibitem [{\citenamefont {Okun}(1982)}]{okun}%
  \BibitemOpen
  \bibfield  {author} {\bibinfo {author} {\bibfnamefont {L.~B.}\ \bibnamefont
  {Okun}},\ }\href@noop {} {\bibfield  {journal} {\bibinfo  {journal} {Sov.
  Phys. JETP},\ }\textbf {\bibinfo {volume} {56}},\ \bibinfo {pages} {502}
  (\bibinfo {year} {1982})}\BibitemShut {NoStop}%
\bibitem [{\citenamefont {Holdom}(1986)}]{holdom}%
  \BibitemOpen
  \bibfield  {author} {\bibinfo {author} {\bibfnamefont {B.}~\bibnamefont
  {Holdom}},\ }\href@noop {} {\bibfield  {journal} {\bibinfo  {journal} {Phys.
  Lett. B},\ }\textbf {\bibinfo {volume} {166}},\ \bibinfo {pages} {196}
  (\bibinfo {year} {1986})}\BibitemShut {NoStop}%
\bibitem [{\citenamefont {Jaeckel}\ and\ \citenamefont
  {Ringwald}(2010)}]{jaeckel_lowenergy}%
  \BibitemOpen
  \bibfield  {author} {\bibinfo {author} {\bibfnamefont {J.}~\bibnamefont
  {Jaeckel}}\ and\ \bibinfo {author} {\bibfnamefont {A.}~\bibnamefont
  {Ringwald}},\ }\Doi {10.1146/annurev.nucl.012809.104433} {\bibfield
  {journal} {\bibinfo  {journal} {Annual Review of Nuclear and Particle
  Science},\ }\textbf {\bibinfo {volume} {60}},\ \bibinfo {pages} {405}
  (\bibinfo {year} {2010})}\BibitemShut {NoStop}%
\bibitem [{\citenamefont {Redondo}\ and\ \citenamefont
  {Ringwald}(2010)}]{redondo_lsw}%
  \BibitemOpen
  \bibfield  {author} {\bibinfo {author} {\bibfnamefont {J.}~\bibnamefont
  {Redondo}}\ and\ \bibinfo {author} {\bibfnamefont {A.}~\bibnamefont
  {Ringwald}},\ }\href@noop {} { (\bibinfo {year} {2010})},\ \bibinfo {note}
  {arXiv:1011.3741 [hep-ph]}\BibitemShut {NoStop}%
\bibitem [{\citenamefont {Van~Bibber}\ \emph {et~al.}(1987)\citenamefont
  {Van~Bibber}, \citenamefont {Dagdeviren}, \citenamefont {Koonin},
  \citenamefont {Kerman},\ and\ \citenamefont {Nelson}}]{axionexp2}%
  \BibitemOpen
  \bibfield  {author} {\bibinfo {author} {\bibfnamefont {K.}~\bibnamefont
  {Van~Bibber}}, \bibinfo {author} {\bibfnamefont {N.~R.}\ \bibnamefont
  {Dagdeviren}}, \bibinfo {author} {\bibfnamefont {S.~E.}\ \bibnamefont
  {Koonin}}, \bibinfo {author} {\bibfnamefont {A.~K.}\ \bibnamefont {Kerman}},
  \ and\ \bibinfo {author} {\bibfnamefont {H.~N.}\ \bibnamefont {Nelson}},\
  }\Doi {10.1103/PhysRevLett.59.759} {\bibfield  {journal} {\bibinfo  {journal}
  {Phys. Rev. Lett.},\ }\textbf {\bibinfo {volume} {59}},\ \bibinfo {pages}
  {759} (\bibinfo {year} {1987})}\BibitemShut {NoStop}%
\bibitem [{\citenamefont {Hoogeveen}\ and\ \citenamefont
  {Ziegenhagen}(1991)}]{Hoogeveen1}%
  \BibitemOpen
  \bibfield  {author} {\bibinfo {author} {\bibfnamefont {F.}~\bibnamefont
  {Hoogeveen}}\ and\ \bibinfo {author} {\bibfnamefont {T.}~\bibnamefont
  {Ziegenhagen}},\ }\Doi {DOI: 10.1016/0550-3213(91)90528-6} {\bibfield
  {journal} {\bibinfo  {journal} {Nuclear Physics B},\ }\textbf {\bibinfo
  {volume} {358}},\ \bibinfo {pages} {3 } (\bibinfo {year} {1991})}\BibitemShut
  {NoStop}%
\bibitem [{\citenamefont {Sikivie}\ \emph {et~al.}(2007)\citenamefont
  {Sikivie}, \citenamefont {Tanner},\ and\ \citenamefont {van
  Bibber}}]{axionexp3}%
  \BibitemOpen
  \bibfield  {author} {\bibinfo {author} {\bibfnamefont {P.}~\bibnamefont
  {Sikivie}}, \bibinfo {author} {\bibfnamefont {D.~B.}\ \bibnamefont {Tanner}},
  \ and\ \bibinfo {author} {\bibfnamefont {K.}~\bibnamefont {van Bibber}},\
  }\Doi {10.1103/PhysRevLett.98.172002} {\bibfield  {journal} {\bibinfo
  {journal} {Phys. Rev. Lett.},\ }\textbf {\bibinfo {volume} {98}},\ \bibinfo
  {pages} {172002} (\bibinfo {year} {2007})},\ \Eprint
  {http://arxiv.org/abs/hep-ph/0701198} {arXiv:hep-ph/0701198} \BibitemShut
  {NoStop}%
\bibitem [{\citenamefont {Mueller}\ \emph {et~al.}(2009)\citenamefont
  {Mueller}, \citenamefont {Sikivie}, \citenamefont {Tanner},\ and\
  \citenamefont {van Bibber}}]{axionexp4}%
  \BibitemOpen
  \bibfield  {author} {\bibinfo {author} {\bibfnamefont {G.}~\bibnamefont
  {Mueller}}, \bibinfo {author} {\bibfnamefont {P.}~\bibnamefont {Sikivie}},
  \bibinfo {author} {\bibfnamefont {D.~B.}\ \bibnamefont {Tanner}}, \ and\
  \bibinfo {author} {\bibfnamefont {K.}~\bibnamefont {van Bibber}},\ }\Doi
  {10.1103/PhysRevD.80.072004} {\bibfield  {journal} {\bibinfo  {journal}
  {Phys. Rev. D},\ }\textbf {\bibinfo {volume} {80}},\ \bibinfo {pages}
  {072004} (\bibinfo {year} {2009})},\ \Eprint {http://arxiv.org/abs/0907.5387}
  {arXiv:0907.5387 [hep-ph]} \BibitemShut {NoStop}%
\bibitem [{\citenamefont {Jaeckel}\ and\ \citenamefont
  {Ringwald}(2008)}]{jaeckel_cavity}%
  \BibitemOpen
  \bibfield  {author} {\bibinfo {author} {\bibfnamefont {J.}~\bibnamefont
  {Jaeckel}}\ and\ \bibinfo {author} {\bibfnamefont {A.}~\bibnamefont
  {Ringwald}},\ }\Doi {DOI: 10.1016/j.physletb.2007.11.071} {\bibfield
  {journal} {\bibinfo  {journal} {Phys. Lett. B},\ }\textbf {\bibinfo {volume}
  {659}},\ \bibinfo {pages} {509 } (\bibinfo {year} {2008})},\ \Eprint
  {http://arxiv.org/abs/arXiv:0707.2063 [hep-ph]} {arXiv:0707.2063 [hep-ph]}
  \BibitemShut {NoStop}%
\bibitem [{\citenamefont {Sikivie}(1983)}]{axionastro1}%
  \BibitemOpen
  \bibfield  {author} {\bibinfo {author} {\bibfnamefont {P.}~\bibnamefont
  {Sikivie}},\ }\Doi {10.1103/PhysRevLett.51.1415} {\bibfield  {journal}
  {\bibinfo  {journal} {Phys. Rev. Lett.},\ }\textbf {\bibinfo {volume} {51}},\
  \bibinfo {pages} {1415} (\bibinfo {year} {1983})}\BibitemShut {NoStop}%
\bibitem [{\citenamefont {Hoogeveen}(1992)}]{Hoogeveen2}%
  \BibitemOpen
  \bibfield  {author} {\bibinfo {author} {\bibfnamefont {F.}~\bibnamefont
  {Hoogeveen}},\ }\Doi {DOI: 10.1016/0370-2693(92)91977-H} {\bibfield
  {journal} {\bibinfo  {journal} {Physics Letters B},\ }\textbf {\bibinfo
  {volume} {288}},\ \bibinfo {pages} {195 } (\bibinfo {year}
  {1992})}\BibitemShut {NoStop}%
\bibitem [{\citenamefont {Cameron}\ \emph {et~al.}(1993)\citenamefont
  {Cameron}, \citenamefont {Cantatore}, \citenamefont {Melissinos},
  \citenamefont {Ruoso}, \citenamefont {Semertzidis}, \citenamefont {Halama},
  \citenamefont {Lazarus}, \citenamefont {Prodell}, \citenamefont {Nezrick},
  \citenamefont {Rizzo},\ and\ \citenamefont {Zavattini}}]{BFRT1}%
  \BibitemOpen
  \bibfield  {author} {\bibinfo {author} {\bibfnamefont {R.}~\bibnamefont
  {Cameron}}, \bibinfo {author} {\bibfnamefont {G.}~\bibnamefont {Cantatore}},
  \bibinfo {author} {\bibfnamefont {A.~C.}\ \bibnamefont {Melissinos}},
  \bibinfo {author} {\bibfnamefont {G.}~\bibnamefont {Ruoso}}, \bibinfo
  {author} {\bibfnamefont {Y.}~\bibnamefont {Semertzidis}}, \bibinfo {author}
  {\bibfnamefont {H.~J.}\ \bibnamefont {Halama}}, \bibinfo {author}
  {\bibfnamefont {D.~M.}\ \bibnamefont {Lazarus}}, \bibinfo {author}
  {\bibfnamefont {A.~G.}\ \bibnamefont {Prodell}}, \bibinfo {author}
  {\bibfnamefont {F.}~\bibnamefont {Nezrick}}, \bibinfo {author} {\bibfnamefont
  {C.}~\bibnamefont {Rizzo}}, \ and\ \bibinfo {author} {\bibfnamefont
  {E.}~\bibnamefont {Zavattini}} (\bibinfo {collaboration} {BFRT
  Collaboration}),\ }\Doi {10.1103/PhysRevD.47.3707} {\bibfield  {journal}
  {\bibinfo  {journal} {Phys. Rev. D},\ }\textbf {\bibinfo {volume} {47}},\
  \bibinfo {pages} {3707} (\bibinfo {year} {1993})}\BibitemShut {NoStop}%
\bibitem [{\citenamefont {Ruoso}\ \emph {et~al.}(1992)\citenamefont {Ruoso},
  \citenamefont {Cameron}, \citenamefont {Cantatore}, \citenamefont
  {Melissinos}, \citenamefont {Semertzidis}, \citenamefont {Halama},
  \citenamefont {Lazarus}, \citenamefont {Prodell}, \citenamefont {Nezrick},
  \citenamefont {Rizzo},\ and\ \citenamefont {Zavattini}}]{BFRT2}%
  \BibitemOpen
  \bibfield  {author} {\bibinfo {author} {\bibfnamefont {G.}~\bibnamefont
  {Ruoso}}, \bibinfo {author} {\bibfnamefont {R.}~\bibnamefont {Cameron}},
  \bibinfo {author} {\bibfnamefont {G.}~\bibnamefont {Cantatore}}, \bibinfo
  {author} {\bibfnamefont {A.~C.}\ \bibnamefont {Melissinos}}, \bibinfo
  {author} {\bibfnamefont {Y.}~\bibnamefont {Semertzidis}}, \bibinfo {author}
  {\bibfnamefont {H.~J.}\ \bibnamefont {Halama}}, \bibinfo {author}
  {\bibfnamefont {D.~M.}\ \bibnamefont {Lazarus}}, \bibinfo {author}
  {\bibfnamefont {A.~G.}\ \bibnamefont {Prodell}}, \bibinfo {author}
  {\bibfnamefont {F.}~\bibnamefont {Nezrick}}, \bibinfo {author} {\bibfnamefont
  {C.}~\bibnamefont {Rizzo}}, \ and\ \bibinfo {author} {\bibfnamefont
  {E.}~\bibnamefont {Zavattini}} (\bibinfo {collaboration} {BFRT
  Collaboration}),\ }\Doi {10.1007/BF01474722} {\bibfield  {journal} {\bibinfo
  {journal} {Z. Phys. C},\ }\textbf {\bibinfo {volume} {56}},\ \bibinfo {pages}
  {505} (\bibinfo {year} {1992})}\BibitemShut {NoStop}%
\bibitem [{\citenamefont {Robilliard}\ \emph {et~al.}(2007)\citenamefont
  {Robilliard}, \citenamefont {Battesti}, \citenamefont {Fouch\'{e}},
  \citenamefont {Mauchain}, \citenamefont {Sautivet}, \citenamefont
  {Amiranoff},\ and\ \citenamefont {Rizzo}}]{BMV}%
  \BibitemOpen
  \bibfield  {author} {\bibinfo {author} {\bibfnamefont {C.}~\bibnamefont
  {Robilliard}}, \bibinfo {author} {\bibfnamefont {R.}~\bibnamefont
  {Battesti}}, \bibinfo {author} {\bibfnamefont {M.}~\bibnamefont
  {Fouch\'{e}}}, \bibinfo {author} {\bibfnamefont {J.}~\bibnamefont
  {Mauchain}}, \bibinfo {author} {\bibfnamefont {A.-M.}\ \bibnamefont
  {Sautivet}}, \bibinfo {author} {\bibfnamefont {F.}~\bibnamefont {Amiranoff}},
  \ and\ \bibinfo {author} {\bibfnamefont {C.}~\bibnamefont {Rizzo}},\ }\Doi
  {10.1103/PhysRevLett.99.190403} {\bibfield  {journal} {\bibinfo  {journal}
  {Phys. Rev. Lett.},\ }\textbf {\bibinfo {volume} {99}},\ \bibinfo {eid}
  {190403} (\bibinfo {year} {2007})},\ \Eprint
  {http://arxiv.org/abs/arXiv:0707.1296 [hep-ex]} {arXiv:0707.1296 [hep-ex]}
  \BibitemShut {NoStop}%
\bibitem [{\citenamefont {Ahlers}\ \emph {et~al.}(2007)\citenamefont {Ahlers},
  \citenamefont {Gies}, \citenamefont {Jaeckel}, \citenamefont {Redondo},\ and\
  \citenamefont {Ringwald}}]{ahlers_paraphoton}%
  \BibitemOpen
  \bibfield  {author} {\bibinfo {author} {\bibfnamefont {M.}~\bibnamefont
  {Ahlers}}, \bibinfo {author} {\bibfnamefont {H.}~\bibnamefont {Gies}},
  \bibinfo {author} {\bibfnamefont {J.}~\bibnamefont {Jaeckel}}, \bibinfo
  {author} {\bibfnamefont {J.}~\bibnamefont {Redondo}}, \ and\ \bibinfo
  {author} {\bibfnamefont {A.}~\bibnamefont {Ringwald}},\ }\Doi
  {10.1103/PhysRevD.76.115005} {\bibfield  {journal} {\bibinfo  {journal}
  {Phys. Rev. D},\ }\textbf {\bibinfo {volume} {76}},\ \bibinfo {eid} {115005}
  (\bibinfo {year} {2007})},\ \Eprint {http://arxiv.org/abs/arXiv:0706.2836
  [hep-ph]} {arXiv:0706.2836 [hep-ph]} \BibitemShut {NoStop}%
\bibitem [{\citenamefont {Chou}\ \emph {et~al.}(2008)\citenamefont {Chou},
  \citenamefont {Wester}, \citenamefont {Baumbaugh}, \citenamefont {Gustafson},
  \citenamefont {Irizarry-Valle}, \citenamefont {Mazur}, \citenamefont
  {Steffen}, \citenamefont {Tomlin}, \citenamefont {Yang},\ and\ \citenamefont
  {Yoo}}]{GammeV}%
  \BibitemOpen
  \bibfield  {author} {\bibinfo {author} {\bibfnamefont {A.~S.}\ \bibnamefont
  {Chou}}, \bibinfo {author} {\bibfnamefont {W.}~\bibnamefont {Wester}},
  \bibinfo {author} {\bibfnamefont {A.}~\bibnamefont {Baumbaugh}}, \bibinfo
  {author} {\bibfnamefont {H.~R.}\ \bibnamefont {Gustafson}}, \bibinfo {author}
  {\bibfnamefont {Y.}~\bibnamefont {Irizarry-Valle}}, \bibinfo {author}
  {\bibfnamefont {P.~O.}\ \bibnamefont {Mazur}}, \bibinfo {author}
  {\bibfnamefont {J.~H.}\ \bibnamefont {Steffen}}, \bibinfo {author}
  {\bibfnamefont {R.}~\bibnamefont {Tomlin}}, \bibinfo {author} {\bibfnamefont
  {X.}~\bibnamefont {Yang}}, \ and\ \bibinfo {author} {\bibfnamefont
  {J.}~\bibnamefont {Yoo}} (\bibinfo {collaboration} {GammeV Collaboration}),\
  }\Doi {10.1103/PhysRevLett.100.080402} {\bibfield  {journal} {\bibinfo
  {journal} {Phys. Rev. Lett.},\ }\textbf {\bibinfo {volume} {100}},\ \bibinfo
  {eid} {080402} (\bibinfo {year} {2008})},\ \Eprint
  {http://arxiv.org/abs/arXiv:0710.3783 [hep-ex]} {arXiv:0710.3783 [hep-ex]}
  \BibitemShut {NoStop}%
\bibitem [{\citenamefont {Ahlers}\ \emph {et~al.}(2008)\citenamefont {Ahlers},
  \citenamefont {Gies}, \citenamefont {Jaeckel}, \citenamefont {Redondo},\ and\
  \citenamefont {Ringwald}}]{ahlers_laser}%
  \BibitemOpen
  \bibfield  {author} {\bibinfo {author} {\bibfnamefont {M.}~\bibnamefont
  {Ahlers}}, \bibinfo {author} {\bibfnamefont {H.}~\bibnamefont {Gies}},
  \bibinfo {author} {\bibfnamefont {J.}~\bibnamefont {Jaeckel}}, \bibinfo
  {author} {\bibfnamefont {J.}~\bibnamefont {Redondo}}, \ and\ \bibinfo
  {author} {\bibfnamefont {A.}~\bibnamefont {Ringwald}},\ }\Doi
  {10.1103/PhysRevD.77.095001} {\bibfield  {journal} {\bibinfo  {journal}
  {Phys. Rev. D},\ }\textbf {\bibinfo {volume} {77}},\ \bibinfo {eid} {095001}
  (\bibinfo {year} {2008})},\ \Eprint {http://arxiv.org/abs/arXiv:0711.4991
  [hep-ph]} {arXiv:0711.4991 [hep-ph]} \BibitemShut {NoStop}%
\bibitem [{\citenamefont {Afanasev}\ \emph {et~al.}(2008)\citenamefont
  {Afanasev}, \citenamefont {Baker}, \citenamefont {Beard}, \citenamefont
  {Biallas}, \citenamefont {Boyce}, \citenamefont {Minarni}, \citenamefont
  {Ramdon}, \citenamefont {Shinn},\ and\ \citenamefont
  {Slocum}}]{afanasev_2008}%
  \BibitemOpen
  \bibfield  {author} {\bibinfo {author} {\bibfnamefont {A.}~\bibnamefont
  {Afanasev}}, \bibinfo {author} {\bibfnamefont {O.~K.}\ \bibnamefont {Baker}},
  \bibinfo {author} {\bibfnamefont {K.~B.}\ \bibnamefont {Beard}}, \bibinfo
  {author} {\bibfnamefont {G.}~\bibnamefont {Biallas}}, \bibinfo {author}
  {\bibfnamefont {J.}~\bibnamefont {Boyce}}, \bibinfo {author} {\bibfnamefont
  {M.}~\bibnamefont {Minarni}}, \bibinfo {author} {\bibfnamefont
  {R.}~\bibnamefont {Ramdon}}, \bibinfo {author} {\bibfnamefont
  {M.}~\bibnamefont {Shinn}}, \ and\ \bibinfo {author} {\bibfnamefont
  {P.}~\bibnamefont {Slocum}},\ }\Doi {10.1103/PhysRevLett.101.120401}
  {\bibfield  {journal} {\bibinfo  {journal} {Phys. Rev. Lett.},\ }\textbf
  {\bibinfo {volume} {101}},\ \bibinfo {eid} {120401} (\bibinfo {year}
  {2008})},\ \Eprint {http://arxiv.org/abs/arXiv:0806.2631 [hep-ex]}
  {arXiv:0806.2631 [hep-ex]} \BibitemShut {NoStop}%
\bibitem [{\citenamefont {Fouch\'{e}}\ \emph {et~al.}(2008)\citenamefont
  {Fouch\'{e}}, \citenamefont {Robilliard}, \citenamefont {Faure},
  \citenamefont {Rizzo}, \citenamefont {Mauchain}, \citenamefont {Nardone},
  \citenamefont {Battesti}, \citenamefont {Martin}, \citenamefont {Sautivet},
  \citenamefont {Paillard},\ and\ \citenamefont {Amiranoff}}]{fouche}%
  \BibitemOpen
  \bibfield  {author} {\bibinfo {author} {\bibfnamefont {M.}~\bibnamefont
  {Fouch\'{e}}}, \bibinfo {author} {\bibfnamefont {C.}~\bibnamefont
  {Robilliard}}, \bibinfo {author} {\bibfnamefont {S.}~\bibnamefont {Faure}},
  \bibinfo {author} {\bibfnamefont {C.}~\bibnamefont {Rizzo}}, \bibinfo
  {author} {\bibfnamefont {J.}~\bibnamefont {Mauchain}}, \bibinfo {author}
  {\bibfnamefont {M.}~\bibnamefont {Nardone}}, \bibinfo {author} {\bibfnamefont
  {R.}~\bibnamefont {Battesti}}, \bibinfo {author} {\bibfnamefont
  {L.}~\bibnamefont {Martin}}, \bibinfo {author} {\bibfnamefont {A.-M.}\
  \bibnamefont {Sautivet}}, \bibinfo {author} {\bibfnamefont {J.-L.}\
  \bibnamefont {Paillard}}, \ and\ \bibinfo {author} {\bibfnamefont
  {F.}~\bibnamefont {Amiranoff}},\ }\Doi {10.1103/PhysRevD.78.032013}
  {\bibfield  {journal} {\bibinfo  {journal} {Phys. Rev. D},\ }\textbf
  {\bibinfo {volume} {78}},\ \bibinfo {eid} {032013} (\bibinfo {year}
  {2008})},\ \Eprint {http://arxiv.org/abs/arXiv:0808.2800 [hep-ex]}
  {arXiv:0808.2800 [hep-ex]} \BibitemShut {NoStop}%
\bibitem [{\citenamefont {Afanasev}\ \emph {et~al.}(2009)\citenamefont
  {Afanasev}, \citenamefont {Baker}, \citenamefont {Beard}, \citenamefont
  {Biallas}, \citenamefont {Boyce}, \citenamefont {Minarni}, \citenamefont
  {Ramdon}, \citenamefont {Shinn},\ and\ \citenamefont
  {Slocum}}]{afanasev_2009}%
  \BibitemOpen
  \bibfield  {author} {\bibinfo {author} {\bibfnamefont {A.}~\bibnamefont
  {Afanasev}}, \bibinfo {author} {\bibfnamefont {O.}~\bibnamefont {Baker}},
  \bibinfo {author} {\bibfnamefont {K.}~\bibnamefont {Beard}}, \bibinfo
  {author} {\bibfnamefont {G.}~\bibnamefont {Biallas}}, \bibinfo {author}
  {\bibfnamefont {J.}~\bibnamefont {Boyce}}, \bibinfo {author} {\bibfnamefont
  {M.}~\bibnamefont {Minarni}}, \bibinfo {author} {\bibfnamefont
  {R.}~\bibnamefont {Ramdon}}, \bibinfo {author} {\bibfnamefont
  {M.}~\bibnamefont {Shinn}}, \ and\ \bibinfo {author} {\bibfnamefont
  {P.}~\bibnamefont {Slocum}},\ }\Doi {DOI: 10.1016/j.physletb.2009.07.055}
  {\bibfield  {journal} {\bibinfo  {journal} {Phys. Lett. B},\ }\textbf
  {\bibinfo {volume} {679}},\ \bibinfo {pages} {317 } (\bibinfo {year}
  {2009})},\ \Eprint {http://arxiv.org/abs/arXiv:0810.4189 [hep-ex]}
  {arXiv:0810.4189 [hep-ex]} \BibitemShut {NoStop}%
\bibitem [{\citenamefont {Ehret}\ \emph {et~al.}(2009)\citenamefont {Ehret}
  \emph {et~al.}}]{ALPS}%
  \BibitemOpen
  \bibfield  {author} {\bibinfo {author} {\bibfnamefont {K.}~\bibnamefont
  {Ehret}} \emph {et~al.} (\bibinfo {collaboration} {ALPS Collaboration}),\
  }\Doi {DOI: 10.1016/j.nima.2009.10.102} {\bibfield  {journal} {\bibinfo
  {journal} {Nucl. Instrum. Meth. A},\ }\textbf {\bibinfo {volume} {612}},\
  \bibinfo {pages} {83 } (\bibinfo {year} {2009})},\ \Eprint
  {http://arxiv.org/abs/arXiv:0905.4159 [physics.ins-det]} {arXiv:0905.4159
  [physics.ins-det]} \BibitemShut {NoStop}%
\bibitem [{\citenamefont {Ehret}\ \emph {et~al.}(2010)\citenamefont {Ehret}
  \emph {et~al.}}]{ALPS2010}%
  \BibitemOpen
  \bibfield  {author} {\bibinfo {author} {\bibfnamefont {K.}~\bibnamefont
  {Ehret}} \emph {et~al.} (\bibinfo {collaboration} {ALPS Collaboration}),\
  }\href@noop {} {\bibfield  {journal} {\bibinfo  {journal} {\hspace{1pt}}}
  (\bibinfo {year} {2010})},\ \Eprint {http://arxiv.org/abs/1004.1313}
  {arXiv:1004.1313 [hep-ex]} \BibitemShut {NoStop}%
\bibitem [{\citenamefont {Povey}\ \emph {et~al.}(2010)\citenamefont {Povey},
  \citenamefont {Hartnett},\ and\ \citenamefont {Tobar}}]{rp_lsw1}%
  \BibitemOpen
  \bibfield  {author} {\bibinfo {author} {\bibfnamefont {R.~G.}\ \bibnamefont
  {Povey}}, \bibinfo {author} {\bibfnamefont {J.~G.}\ \bibnamefont {Hartnett}},
  \ and\ \bibinfo {author} {\bibfnamefont {M.~E.}\ \bibnamefont {Tobar}},\
  }\Doi {10.1103/PhysRevD.82.052003} {\bibfield  {journal} {\bibinfo  {journal}
  {Phys. Rev. D},\ }\textbf {\bibinfo {volume} {82}},\ \bibinfo {pages}
  {052003} (\bibinfo {year} {2010})},\ \Eprint
  {http://arxiv.org/abs/1003.0964v2} {arXiv:1003.0964v2 [hep-ex]} \BibitemShut
  {NoStop}%
\bibitem [{\citenamefont {Wagner}\ \emph {et~al.}(2010)\citenamefont {Wagner},
  \citenamefont {Rybka}, \citenamefont {Hotz}, \citenamefont {Rosenberg},
  \citenamefont {Asztalos}, \citenamefont {Carosi}, \citenamefont {Hagmann},
  \citenamefont {Kinion}, \citenamefont {van Bibber}, \citenamefont {Hoskins},
  \citenamefont {Martin}, \citenamefont {Sikivie}, \citenamefont {Tanner},
  \citenamefont {Bradley},\ and\ \citenamefont {Clarke}}]{admx_lsw}%
  \BibitemOpen
  \bibfield  {author} {\bibinfo {author} {\bibfnamefont {A.}~\bibnamefont
  {Wagner}}, \bibinfo {author} {\bibfnamefont {G.}~\bibnamefont {Rybka}},
  \bibinfo {author} {\bibfnamefont {M.}~\bibnamefont {Hotz}}, \bibinfo {author}
  {\bibfnamefont {L.~J.}\ \bibnamefont {Rosenberg}}, \bibinfo {author}
  {\bibfnamefont {S.~J.}\ \bibnamefont {Asztalos}}, \bibinfo {author}
  {\bibfnamefont {G.}~\bibnamefont {Carosi}}, \bibinfo {author} {\bibfnamefont
  {C.}~\bibnamefont {Hagmann}}, \bibinfo {author} {\bibfnamefont
  {D.}~\bibnamefont {Kinion}}, \bibinfo {author} {\bibfnamefont
  {K.}~\bibnamefont {van Bibber}}, \bibinfo {author} {\bibfnamefont
  {J.}~\bibnamefont {Hoskins}}, \bibinfo {author} {\bibfnamefont
  {C.}~\bibnamefont {Martin}}, \bibinfo {author} {\bibfnamefont
  {P.}~\bibnamefont {Sikivie}}, \bibinfo {author} {\bibfnamefont {D.~B.}\
  \bibnamefont {Tanner}}, \bibinfo {author} {\bibfnamefont {R.}~\bibnamefont
  {Bradley}}, \ and\ \bibinfo {author} {\bibfnamefont {J.}~\bibnamefont
  {Clarke}},\ }\Doi {10.1103/PhysRevLett.105.171801} {\bibfield  {journal}
  {\bibinfo  {journal} {Phys. Rev. Lett.},\ }\textbf {\bibinfo {volume}
  {105}},\ \bibinfo {pages} {171801} (\bibinfo {year} {2010})},\ \Eprint
  {http://arxiv.org/abs/1007.3766} {arXiv:1007.3766 [hep-ex]} \BibitemShut
  {NoStop}%
\bibitem [{\citenamefont {Belinfante}(1939)}]{belinfante}%
  \BibitemOpen
  \bibfield  {author} {\bibinfo {author} {\bibfnamefont {F.}~\bibnamefont
  {Belinfante}},\ }\href@noop {} {\bibfield  {journal} {\bibinfo  {journal}
  {Physica},\ }\textbf {\bibinfo {volume} {6}},\ \bibinfo {pages} {887}
  (\bibinfo {year} {1939})}\BibitemShut {NoStop}%
\bibitem [{\citenamefont {Pound}(1946)}]{Pound}%
  \BibitemOpen
  \bibfield  {author} {\bibinfo {author} {\bibfnamefont {R.~V.}\ \bibnamefont
  {Pound}},\ }\Doi {10.1063/1.1770414} {\bibfield  {journal} {\bibinfo
  {journal} {Review of Scientific Instruments},\ }\textbf {\bibinfo {volume}
  {17}},\ \bibinfo {pages} {490} (\bibinfo {year} {1946})}\BibitemShut
  {NoStop}%
\bibitem [{\citenamefont {Stein}\ and\ \citenamefont {Turneaure}(1975)}]{Q10}%
  \BibitemOpen
  \bibfield  {author} {\bibinfo {author} {\bibfnamefont {S.~R.}\ \bibnamefont
  {Stein}}\ and\ \bibinfo {author} {\bibfnamefont {J.~P.}\ \bibnamefont
  {Turneaure}},\ }\Doi {10.1109/PROC.1975.9920} {\bibfield  {journal} {\bibinfo
   {journal} {Proceedings of the IEEE},\ }\textbf {\bibinfo {volume} {63}},\
  \bibinfo {pages} {1249 } (\bibinfo {year} {1975})}\BibitemShut {NoStop}%
\bibitem [{\citenamefont {Ivanov}\ and\ \citenamefont {Tobar}(2006)}]{eugene1}%
  \BibitemOpen
  \bibfield  {author} {\bibinfo {author} {\bibfnamefont {E.~N.}\ \bibnamefont
  {Ivanov}}\ and\ \bibinfo {author} {\bibfnamefont {M.~E.}\ \bibnamefont
  {Tobar}},\ }\Doi {10.1109/TMTT.2006.879172} {\bibfield  {journal} {\bibinfo
  {journal} {Microwave Theory and Techniques, IEEE Transactions on},\ }\textbf
  {\bibinfo {volume} {54}},\ \bibinfo {pages} {3284 } (\bibinfo {year}
  {2006})}\BibitemShut {NoStop}%
\bibitem [{\citenamefont {Santiago}\ \emph {et~al.}(1996)\citenamefont
  {Santiago}, \citenamefont {Dick},\ and\ \citenamefont {Wang}}]{dick_lsf}%
  \BibitemOpen
  \bibfield  {author} {\bibinfo {author} {\bibfnamefont {D.}~\bibnamefont
  {Santiago}}, \bibinfo {author} {\bibfnamefont {G.}~\bibnamefont {Dick}}, \
  and\ \bibinfo {author} {\bibfnamefont {R.}~\bibnamefont {Wang}},\ }in\ \Doi
  {10.1109/FREQ.1996.560254} {\emph {\bibinfo {booktitle} {Frequency Control
  Symposium, 1996. 50th., Proceedings of the 1996 IEEE International.}}}\
  (\bibinfo {year} {1996})\ pp.\ \bibinfo {pages} {772 --775}\BibitemShut
  {NoStop}%
\end{thebibliography}
\end{document}